# Role of polarized tip-enhanced Raman spectroscopy in the enhancement of interface optical phonon modes in AlGaN multi-quantum wells

A.K. Sivadasan[1†*], Prajit Dhara[2], Chirantan Singha[3], Raktima Basu[1*], Santanu Parida[1], A. Bhattacharyya[3], Kishore K. Madapu[1], and Sandip Dhara[1*]

[1] Surface and Nanoscience Division, Indira Gandhi Centre for Atomic Research, Homi Bhabha National Institute, Kalpakkam-603102, India

[2] Department of Electrical and Electronics Engineering, Birla Institute of Technology and Science-Pilani, Pilani - 333031, India

[3] Centre for Research in Nanoscience and Nanotechnology, University of Calcutta, JD2, Sector–III, Saltlake City, Kolkata–700106, West Bengal, India

[†] Presently at Govt. Higher Sec. School, Cheruthuruthy- 679531, Kerala, India

[*] Corresponding authors

A.K. Sivadasan (sivankondazhy@gmail.com)

Raktima Basu (raktimabasu14@gmail.com)

Sandip Dhara  (dhara@igcar.gov.in)

**Abstract**

Group III nitride based two-dimensional multi-quantum well (MQW) nanostructures find remarkable applications in the visible to ultraviolet light sources. The interface optical (IFO) phonon modes in a *c*-axis oriented superlattice of [Al$_{0.35}$Ga$_{0.65}$N (~1.75 nm)/Al$_{0.55}$Ga$_{0.45}$N (~2nm)]$_{20}$ MQWs are observed using tip-enhanced Raman spectroscopic (TERS) studies. The near-field studies using TERS probe with an Au spherical nanoparticle of ~ 200 nm diameter were carried out at ambient conditions showing approximately two to three orders of enhancement in the Raman intensities.  The interface phonon mode belonging to $E_1$ symmetry [IFO($E_1$)] vibrating normal to the *c*-axis of MQWs appeared to be more prominent in the case of TERS measurement compared to that for the other interface phonon mode of $A_1$ symmetry.

The confined electric field of the polarized electro-magnetic excitation using TERS probe, parallel to the plane of the interface of MQW, is made responsible for the plasmonic enhancement of IFO($E_1$) phonon mode. The confinement was verified using finite-difference time-domain simulation.

1. **Introduction**

Quantum wells (QW) are two-dimensional (2D) nanostructures. They are well known for the unidirectional phonon and carrier confinement effects, as well as for the bidirectional transport properties. The major components of various optoelectronic devices of III-nitrides are fabricated using superlattices or in other words 'lattice of lattices' structures. The assembly of a superlattice exhibits a substantial penetration of wave function between the neighboring layers because of the tunneling effects leading to the formation of a multi-quantum well (MQW) [1,2]. Generally, the *c*-plane oriented AlGaN/Al(Ga)N heterostructures of such MQW or superlattice generate some stationary charges at the hetero-interfaces due to the discontinuities of spontaneous polarization ($P^{SP}$) and piezoelectric polarization ($P^{PZ}$) between successive layers. These polarization effects can generate a 2D-electron gas (2DEG) at the interfaces, which can be beneficial for the development of various optoelectronic devices [3-5]. The capability to sustain a 2DEG at the hetero-junctions makes the III-nitrides MQWs or superlattices especially AlGaN/Al(Ga)N with various Al and Ga percentage suitable for the fabrication of high electron mobility transistors along with the hetero-junction field effect transistors and bipolar junction transistors [3,4,6,7]. The superlattices and MQWs of AlGaN/Al(Ga)N with different concentrations of its constituent elements are used as an important component in different optoelectronic devices such as blue, ultra-violet (UV) and deep-UV light emitting diodes and laser diodes as a buffer layer, active region. They are also used to form an electron blocking layers by confining carrier to improve their internal and external quantum efficiency [7-9].

The vibration of surface atoms with their amplitudes confined close to the surface is known as surface optical phonon mode, and its strength decrease exponentially with depth. The effect of electron-phonon scattering originating from the surface modulation is the major reason for the pronounced generation of surface optical phonon modes. The surface optical phonon modes becomes prominent with the breakdown of translational symmetry of the surface potential because of the presence of periodic textures engraved on the surface of nanostructures [10,11]. Similarly, in the case of MQWs or superlattices interface optical (IFO) phonon modes can be observed because of the vibration of atoms at the interfaces of heterojunctions. The breakdown of translational symmetry of surface potential pertaining to the periodicity of superlattice structures leads to the generation of an additional crystal momentum, which relaxes the selection rules of Raman scattering and leads to the observation of non-zone center phonons (wavevector, $q \neq 0$) in the spectra. The additional momenta leading to the creation of non-zone center interface optical phonons can be absorbed by the periodic structure of the superlattices as quantized units of $2\pi/\lambda$, where $\lambda$ is the periodicity of the superlattices or MQWs [12]. It is also well known that the frequency of surface optical or interface optical phonon mode depends on the dielectric constant of the surrounding medium ($\varepsilon_m$) as well as the shape and size of the nanostructures [10-13].

Tip-enhanced Raman spectroscopy (TERS) is a non-destructive optical characterization technique exclusively meant for the study of nanostructures in the sub-wavelength regime [14,15]. The TERS study mainly utilizes the advantages of optical confinement by means of plasmonics. The coherent oscillations of surface plasmons with respect to the excitation may lead to the localization of light bounded to the surface of plasmonic metal nanostructures. The confined or concentrated light generated in the vicinity of such metallic nanostructures, especially Au and Ag, enables enhanced light-matter interactions [16-19]. The major reports based on TERS studies mainly focused on covalently bonded materials such as bio-organic or carbon molecules with an enhancement factor (*EF*) of eight orders [14,15]. The value of the fractional ionic character of AlN based compounds is very close to unity as in the case of pure ionic crystallites. There are only a few reports available on the TERS based studies of such compounds with a significantly high fractional ionic character with significantly low enhancement factor. The 2DEG in AlGaN MQWs is generated at the interface of the superlattice because of the polarization discontinuity. The 2DEG are confined in one direction and free to move along other two directions. Hence, there is a high probability of influencing the 2DEG by the confined electric field. Since in TERS, the electric field is confined with high spatial resolution, the behavior of the 2DEG and other phenomena arising at the interfaces will be highly influenced. Moreover, the electron-phonon interaction also will be modified at the interface. Therefore, the present study aims to study the influence of TERS tip assisted highly confined electric field on the interface phonon modes.

The present study is mainly focused on the TERS studies of 2D MQWs of AlGaN with a higher degree of ionic character than the covalent counterpart. The TERS probe attached with plasmonic Au nanoparticle is also used to confine the electric field of the electro-magnetic excitation parallel to the plane of the interface of MQW. So, the present study also implemented polarized TERS as a localized probe to study its influence on various interface optical phonon modes and understanding the role of confinement of electric field generated at the consecutive layers of plasma-assisted molecular beam epitaxial (PAMBE) grown $[Al_{0.35}Ga_{0.65}N$ (~1.75 nm)/$Al_{0.55}Ga_{0.45}N$ (~2nm)$]_{20}$ MQWs. The finite-difference time-domain (FDTD) simulation is carried out to understand the confinement as well as the interaction of the electric field with the sample.

2. **Materials and Methods**

    *2.1 Growth and structural characterization of multi-quantum well*

The MQW of $[Al_{0.35}Ga_{0.65}N$ (~1.75 nm)/$Al_{0.55}Ga_{0.45}N$ (~2nm)$]_{20}$ on AlGaN (~300 nm)/AlN (~200 nm) buffer layer was grown on a *c*-plane oriented sapphire substrate using PAMBE (VEECO; Gen 930) system. Standard effusion cells for group III sources were used along with a radio-frequency (RF) plasma source for N activation. The detailed growth mechanism of MQW is discussed in our previous reports [10,13]. In brief, high purity Al (7N) and Ga (6N) metals in two zone effusion cells were used to generate their fluxes and $N_2$ gas (6N), assisted by RF plasma source, was used as an activator. The growth process was executed in three steps, namely, nitridation by formation of nitrogen plasma (power: 350 W, flow rate: ~1.7 sccm) to convert the top layer of the sapphire ($Al_2O_3$) to AlN at a temperature of 1023K; followed by the formation of AlN buffer layer in the presence of excess Al flux at 1023K in the high vacuum ($3.67 \times 10^{-7}$ mbar). Finally, AlGaN film was deposited with Al flux of $2.17 \times 10^{-7}$ mbar, and Ga flux of $6.83 \times 10^{-7}$ mbar, at 1043K. The AlN islands grown on the substrate were utilized as nucleation centers for the formation of AlGaN films. Transmission electron microscopy (TEM; Tecnai 20 G20) operating at 200 keV was utilized for cross-sectional microstructural studies of $[Al_{0.35}Ga_{0.65}N/Al_{0.55}Ga_{0.45}N]_{20}$ superlattices.

*2.2 Tip-enhanced Raman spectroscopic charaterization*

The schematic of the TERS technique used for the investigation of interface optical phonon modes of 2D MQW nanostructures are shown in supplementary Figure S1. The generation of evanescent waves and optical confinement is achieved with the aid of plasmonics by means of Au nanoparticle (NP; ~200 nm) attached to the TERS probe. TERS studies were carried out in the backscattering configuration using the Raman spectrometer (inVia, Renishaw, UK) which was coupled to the scanning probe microscope (SPM) (MultiView 4000; Nanonics, Israel). The Raman scattering data were recorded using vertically polarized (400:1) 514.5 nm excitation on MQWs with an integration time of 10 s. The scattered signals were dispersed with 1800 gr·mm$^{-1}$ grating and recorded with a thermoelectrically cooled CCD detector. The far-field (excluding the TERS probe) and near-field (including the TERS probe) spectra were collected through the same 50× objective with N.A. value of 0.45 leading to focused spot size, $d \sim 1.5$ μm. A nominal laser power of ~0.6 mW with a power density of 34 KW·cm$^{-2}$ was used for the study

*2.3 Finite-difference time-domain simulation details*

Commercial 3D FDTD numerical simulation software (Lumerical FDTD Solutions, Canada) was used to understand the localized field distribution around the coupled system of the Au tip and the analyte. The TERS tip was configured by keeping Au particle with 200 nm diameter on a tilted conical shaped glass tip. The plasmonic probe was simulated considering confinement of 514.5 nm excitation light source at the Au NP using perfectly matched–layer boundary conditions. A plane wave source polarized in the *X*-axis direction was used to inject the field along the *Z*-axis for the excitation of the coupled Au tip and MQW system. The simulation time was tuned to 1000 fs to allow the energy field to decay completely. During all the numerical simulations, the mesh size was kept as $2 \times 2 \times 1$ nm. Further, the tip was positioned 10 nm above the sample surface to simulate the approximate gap expected during a typical scan. The near-field intensity maps for the multilayer were simulated and evaluated with the experimental data.

## 3. Results and Discussion

### 3.1 Morphology

The schematic structure of $[Al_{0.35}Ga_{0.65}N$ (~1.75 nm)/$Al_{0.55}Ga_{0.45}N$ (~2nm)$]_{20}$ (Fig. 1(a)), as well as, a transmission electron microscopic (TEM) micrograph (Fig. 1(b)) of MQWs are shown. The periodic bi-layers of 3.75 nm are measured for the consecutive layers of the superlattice. Hence, the total thickness of the MQW is 75 nm.

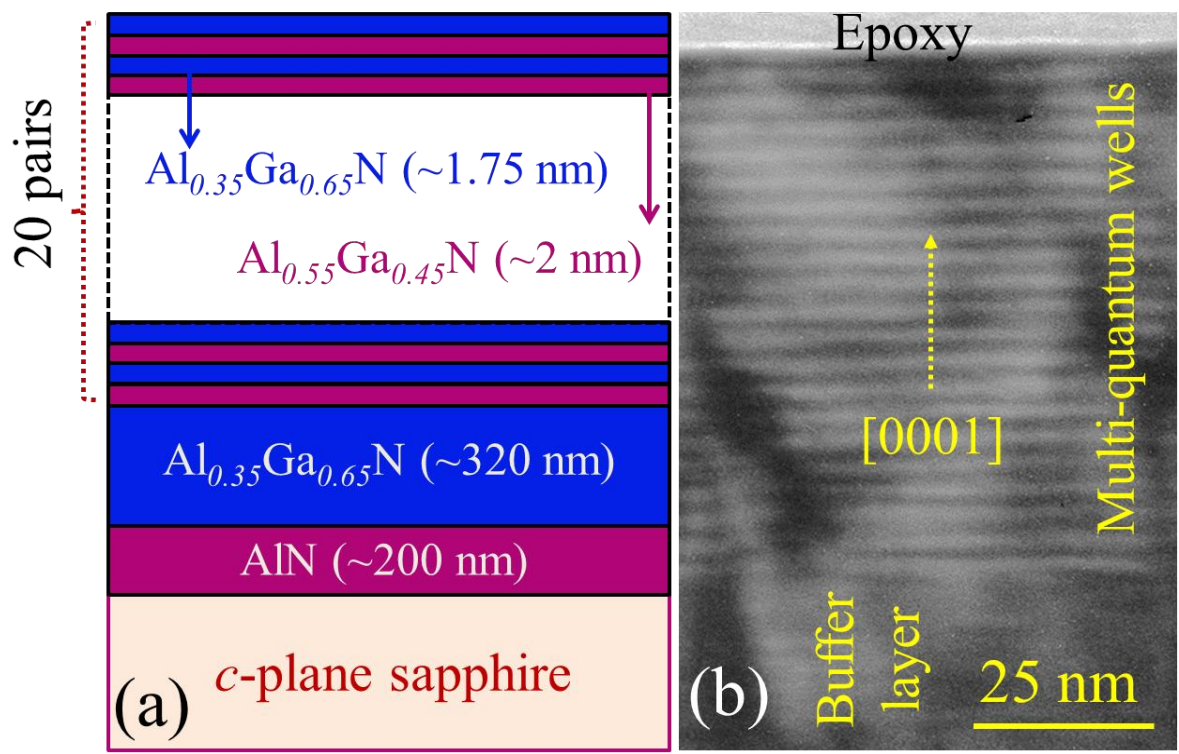

**Figure 1.** (a) The schematic structure and (b) TEM micrograph of [$Al_{0.35}Ga_{0.65}N$ (~1.75 nm)/$Al_{0.55}Ga_{0.45}N$ (~2nm)]$_{20}$ MQWs

*3.2 FDTD simulation*

In order to examine the localized field enhancement in the presence of TERS tip on the multilayer, the near-field electric field intensity imaging for the coupled system was simulated and analyzed using FDTD numerical simulations. The schematic of the simulated structures (Fig. 2a) is shown along with the close view for the projection in *XZ*–plane (Fig. 2b). The near-field distribution projected in *XY* plane shows that the electric field of the polarized source is concentrated along the *X*-axis (Fig. 2c). The near-field electric field intensity distribution along *XZ*–plane for the coupled system shows that the field is confined ~ 100 nm in the *Z* direction, covering all the interfaces of the MQW with 75 nm of thickness (Fig. 2d).

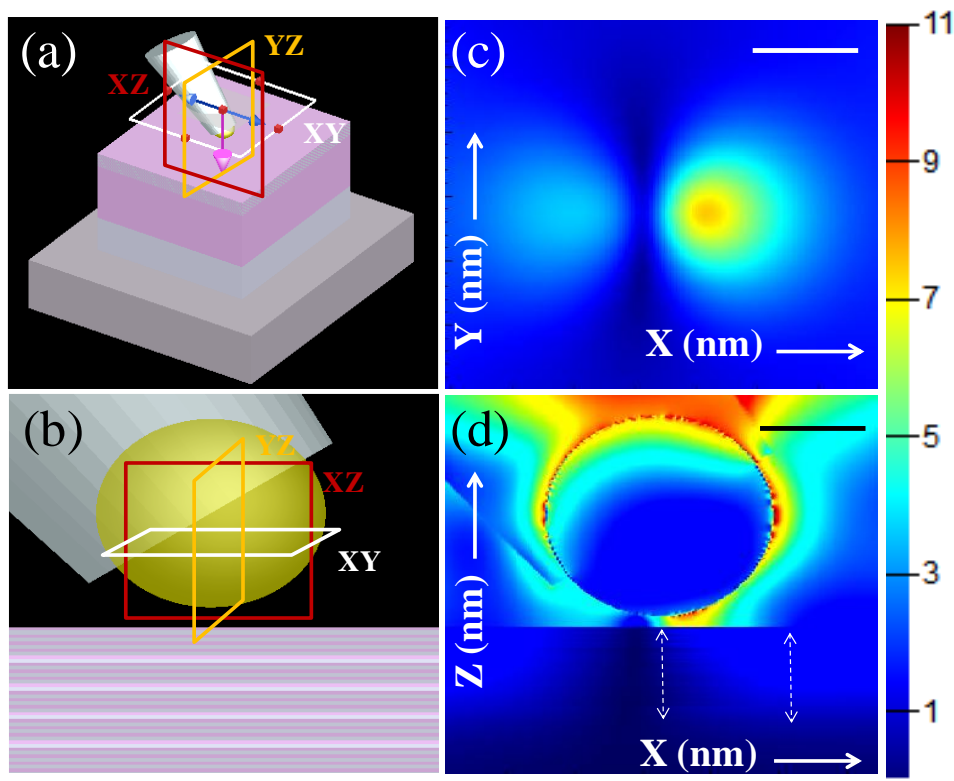

**Figure 2.** The schematics of (a) the simulated structure along with (b) the zoomed view for the projection in *XZ*–plane for the multilayer coupled with TERS tip. The near-field electric field intensity distribution images in (c) *XY* and (d) *XZ*–planes are also shown. Scale bars are 100 nm. The color bars indicate the relative strength of the field. Field penetration along Z direction (depth of the MQW) is around 100 nm, as indicated by dotted double arrows.

*3.3 Far- and near-field Raman spectroscopic studies*

The vibrational properties of the samples with MQW structures of *c*-plane oriented $[Al_{0.35}Ga_{0.65}N/Al_{0.55}Ga_{0.45}N]_{20}$ is shown in Figure 3a. The far-field and near-field measurements were carried out at room temperature (300K) and plotted in the same graph. The observed near- and far-field Raman spectra, shown in Figures 3b and 3c, respectively, are de-convoluted with a Lorentzian function for analyzing the nature and different peak positions of phonon modes. The peaks centered at 578, 660 and 885 cm$^{-1}$ are symmetry allowed GaN-$E_2^H$, AlN-$E_2^H$ and AlN-$A_1$(LO) Raman modes of wurtzite phase [13,20], respectively. The mode at 750 cm$^{-1}$ corresponds to the sapphire substrate. The phonon mode belonging to the substrate is observed, as the MQW sample made of wide band-gap materials (band gap >3.47 eV at 300K) are transparent to 2.41 eV (514.5 nm) excitation. Incidentally, the electric field is observed to be confined in the MQW as shown in the FDTD simulation for *XY* (Figs. 2c) and *XZ* planes (Figs. 2d). It is well reported, both theoretically and experimentally, that the observed phonon modes in between LO$|_{q=0}$ and TO$|_{q=0}$ modes for the MQW structures of AlGaN can be assigned as interface optical phonon modes [13]. The periodic surface modulation and interfaces of superlattice structures of 2D MQWs may lead to the relaxation of selection rules pertaining to Raman scattering owing to the breakdown of translational symmetry of the surface potential [10-13]. Therefore, the broad peak at 802 and 848 cm$^{-1}$ (Figs. 3b and 3c) are assigned as IFO($A_1$) with $A_1$ symmetry and IFO($E_1$) with $E_1$ symmetry, respectively.[13] As discussed earlier, the interfaces of MQWs and buffer layers may absorb an additional momentum as quantized units of interface optical phonon wave vector *q* (=*2π/λ*), where *λ* is the periodicity of the superlattice.[12,13] Broadening of phonon mode, particularly for the interface modes both in far- and near-field measurements, may be due to luminescence in MQW.

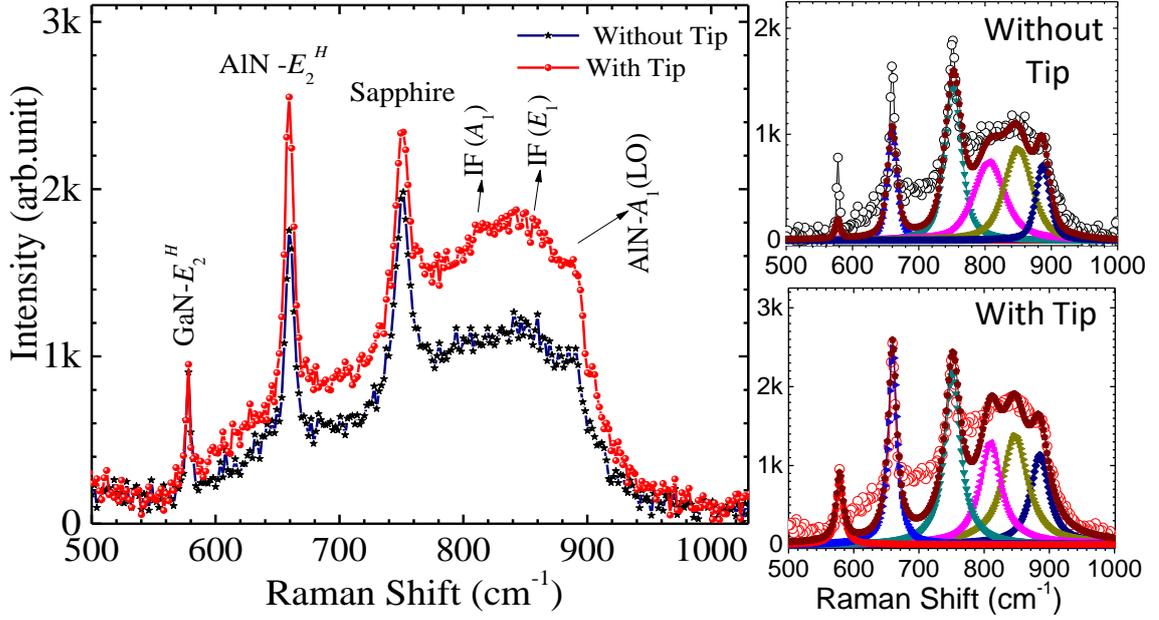

**Figure 3.** (a) Far- and near-field Raman spectra recorded at 300K for AlGaN MQWs without and with the presence of TERS probe, respectively. Deconvoluted spectra correspond to different Raman modes (b) without (far-field) and (c) with (near-field) TERS probe.

The Raman enhancement factor for the recorded TERS spectra is estimated using the relation [21,22],

$$EF = \left( \frac{I_{with-tip}}{I_{without-tip}} - 1 \right) \frac{V_{FF}}{V_{NF}} \quad\text{------------------------------------------(1)}$$

where $I_{with-tip}$ and $I_{without-tip}$ are the recorded Raman scattering intensities with and without the presence of TERS probe in the near- and the far-field, respectively. $V_{FF}$ is the laser excitation interaction volume of the far-field configuration and $V_{NF}$ is the effective plasmonic interaction volume constituted by the near-field evanescent waves. We assume a interaction volume of cylindrical shape (Supplementary informartion Fig. S1) for 2D MQW as $V = \pi(d/2)^2 \delta$, where, $\delta$ is the penetration depth of laser excitation through the MQWs and $d/2$ is the radius of the cylindrical interaction volume. $V_{FF}$ can be calculated as $\sim 971 \times 10^6$ nm$^3$ by considering $d/2 \sim$ 750 nm and $\delta \sim$ 550 nm (thicknesses of buffer layers and MQWs) in case of far-field Raman spectra. Whereas, the presence of TERS probe may confine the laser light and concentrate the field strength into a spot with a diameter ~100 nm (~size of Au NP) for near-field Raman spectral acquisition. With the assumption that evanescent waves can penetrate a semiconducting sample up to a maximum distance of the order of a half-wavelength and considering our FDTD calculation for the specific probe and material structure to a value of 100 nm ; $V_{NF}$ is estimated ~ $78.5\times10^4$ nm$^3$. The typical value of $EF$ is calculated as ~743, which is apprixamately three orders high for the AlN-$E_2^H$ mode. The value of $EF$ for the IFO($A_1$) and IFO($E_1$) modes are calculated to be ~246 and ~695, respectively, which are also two to three orders high. Necessary data may be found in the supplementary information showing detailed calculations and tabulated in Table 1.

**Table 1.** Parameters extracted from fitted Raman spectra with (Fig. 3b) and without (Fig. 3c) TERS tip and the corresponding $EF$ values (supplementary information showing detailed calculation)

| Modes | Intensity (Without tip) | Intensity (With tip) | $EF$ |
|---|---|---|---|
| AlN-$E_2^H$ | 1382 | 2212 | 743 |
| IFO($A_1$) | 766 | 918 | 246 |
| IFO($E_1$) | 735 | 1148 | 695 |

Similar $EF$ values of $E_2^H$ and IFO($A_1$) modes and a higher value of three orders of $EF$ for IFO($E_1$) mode are also recorded for another measurement at a different spot of the sample (supplementary materials Fig. S2 and Table S1). The $EF$ values for IFO($E_1$) mode is always found higher than those calculated for IFO($A_1$) mode. The $EF$ values estimated for the semiconductor AlGaN MQWs, with a higher degree of ionic nature, is observed as low compared to that of reported values (~$10^8$) of covalently bonded materials. Generally, the Raman scattering intensity is proportional to the fluctuations in electric susceptibility, which is the intrinsic property of a material [22-24]. The changes in electric susceptibility with excitation laser are very high for the covalently bonded molecules compared to that of ionically bonded ones [22]. Therefore, the Raman scattering cross-section (scattering efficiency) and hence the *EF* for the $[Al_{0.35}Ga_{0.65}N/Al_{0.55}Ga_{0.45}N]_{20}$ MQWs with higher fractional ionic character show comparatively lower values. However, the interface optical phonon modes along with other symmetry allowed modes are more prominent (Fig. 3) in the case of TERS measurement compared to that of far-field spectra because of the plasmonic confinement of light and electric field enhancement effects. Generally, the interface optical phonon modes may appear with large FWHM and huge luminescence background due to the presence of high electron density or interface charges generated due to the polarization discontinuity in the multi-layers. The confinement of light using TERS tip can also influence the enhancement of these background luminescence of both IFO($E_1$) and IFO($A_1$) phonon modes in the spectra. At the same time, the vibration of surface atoms normal to the *c*-axis, belonging to $E_1$ symmetry (supplementary Fig. S3), evolves and gains prominence in the case of TERS spectra compared to that of the other interface phonon mode of $A_1$ symmetry (Fig. 3). The confined electric field of electro-magnetic excitation using $Ar^+$ laser normal to the long axis of TERS probe aligns parallel to the plane of the film (FDTD simulation results in Fig. 2c) and is made responsible for the maximum plasmonic enhancement of IFO($E_1$) phonon mode. High *EF* value for $E_2^H$ mode, comparable to that for the IFO($E_1$) mode, also can be understood as in this case also the atoms vibrate normal to the *c*-axis (supplementary Fig. S3), which is also the direction of polarization induced and confined by the TERS tip.

4. **Conclusion**

In conclusion, interface optical (IFO) phonon modes along with symmetry allowed modes are observed in the near-field tip-enhanced Raman spectroscopy (TERS) spectra of *c*-axis oriented $[Al_{0.35}Ga_{0.65}N$ (~1.75 nm)/$Al_{0.55}Ga_{0.45}N$ (~2nm)$]_{20}$ multi-quantum well (MQW) structures recorded at ambient conditions. The observation of the interface optical phonon modes is ascribed to the relaxation of selection rules of Raman scattering owing to the breakdown of translational symmetry of the surface potential in the presence of the periodic interfaces of the superlattice structure of MQWs. Polarized TERS study showed intense Raman spectra of MQWs of III nitrides with a sufficiently high enhancement factor of two to three orders. In the case of interface optical phonon modes, the mode with $E_1$ symmetry which is generated due to the vibration of surface atoms normal to the *c*-axis of the MQWs, appears to be more distinct. The large TERS enhancement of IFO($E_1$) as compared to that for the other interface Raman mode of $A_1$ symmetry is attributed to the presence of a confined electric field parallel to the plane of a superlattice of MQWs, as also confirmed from the finite-difference time-domain simulations.

**Supplementary Materials** 
**Role of polarized tip-enhanced Raman spectroscopy in the enhancement of interface optical phonon modes in AlGaN multi-quantum wells**

A.K. Sivadasan[1,†,*], Prajit Dhara[2], Chirantan Singha[3], Raktima Basu[1,*], Santanu Parida[1], A. Bhattacharyya[3], Kishore K. Madapu[1], and Sandip Dhara[1,*]

[1] Surface and Nanoscience Division, Indira Gandhi Centre for Atomic Research, Homi Bhabha National Institute, Kalpakkam-603102, India

[2] Department of Electrical and Electronics Engineering, Birla Institute of Technology and Science-Pilani, Pilani - 333031, India

[3] Centre for Research in Nanoscience and Nanotechnology, University of Calcutta, JD2, Sector–III, Saltlake City, Kolkata–700106, West Bengal, India

[†] Presently at Govt. Higher Sec. School, Cheruthuruthy- 679531, Kerala, India

[*] Authors to whom correspondence should be addressed:

sivankondazhy@gmail.com; raktimabasu14@gmail.com; madupu@igcar.gov.in

**Calculation of the enhancement factor (*EF*):**

The *EF* of the spectra can be calculated using the Eqn (1). Let us consider $V_{FF}$ is the interaction volume of the far–field laser probe and $V_{NF}$ is the effective interaction volume of near–field TERS probe. We consider a cylindrical interaction volume for 2D MQW as $V = \pi (d/2)^2 \delta$ (Fig. S1); where, $\delta$ is the depth of penetration. The far field spectra were collected using a 50× objective with numerical aperture (N.A.) value of 0.45 (Fig. S2). The corresponding focused spot size of laser beam ($d = 1.22 \lambda/\text{N.A.}$) for an excitation wavelength ($\lambda$) of 514.5 nm could be considered as ~ 1.5 μm. In the case of near–field spectra (Fig. S2), the effective electric field can be considered as it is concentrated down to a spot size of $d$~200 nm due to Au plasmonic confinement.

Therefore; $V_{FF} = \pi \times 750^2 \text{ nm}^2 \times 550 \text{ nm}$ ( $\delta$ = Thickness of buffer + MQW) = $971.45 \times 10^6$ nm$^3$

$V_{NF} = \pi \times 50^2 \text{ nm}^2 \times 100 \text{ nm}$ ($\delta$ ~ plasmonic interaction volume as estimated in the FDTD calculation) = $78.5 \times 10^4$ nm$^3$

$V_{FF}/V_{NF} = (971.45/78.5) \times 10^2 = 1237.5$

For example, the *EF* for AlN-$E_2^H$ mode; $EF(AlN - E_2^H) = \left(\frac{2211.6}{1381.6} - 1\right)\frac{V_{FF}}{V_{NF}}$

$$= 0.6 \times 1237.5 \approx 743$$

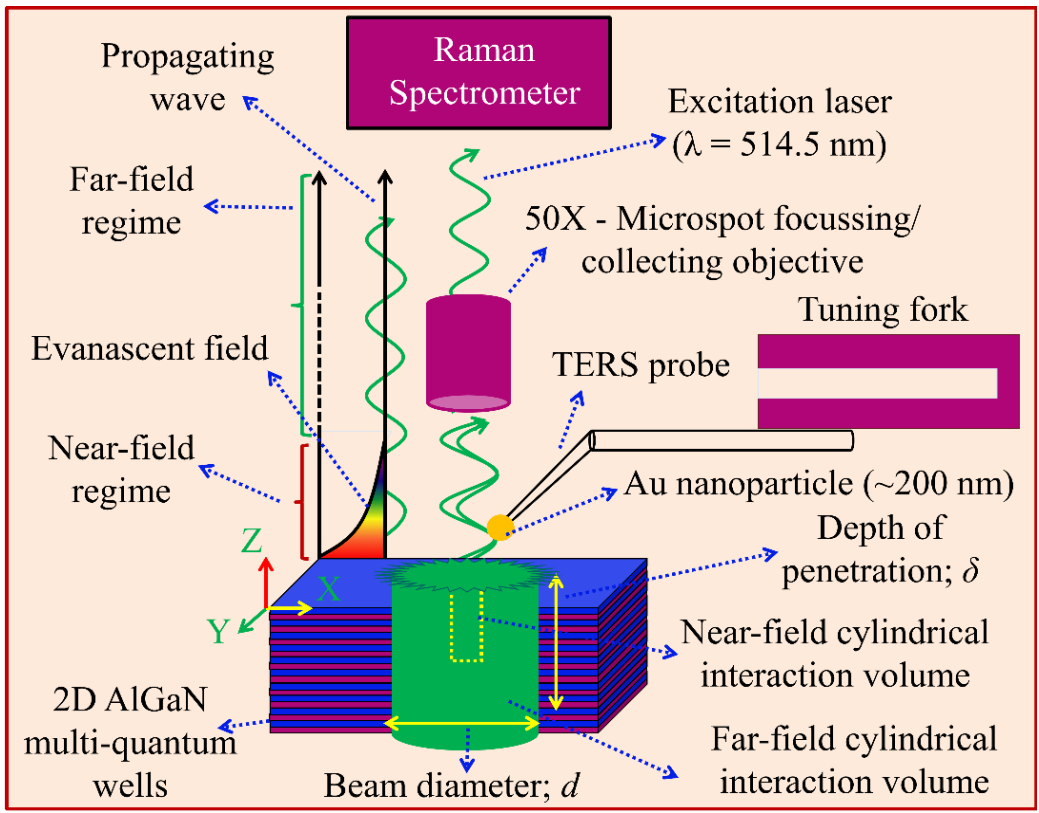

FIG S1. The schematic sketch of the experimental set up used for the near-field TERS measurements of AlGaN MQWs.

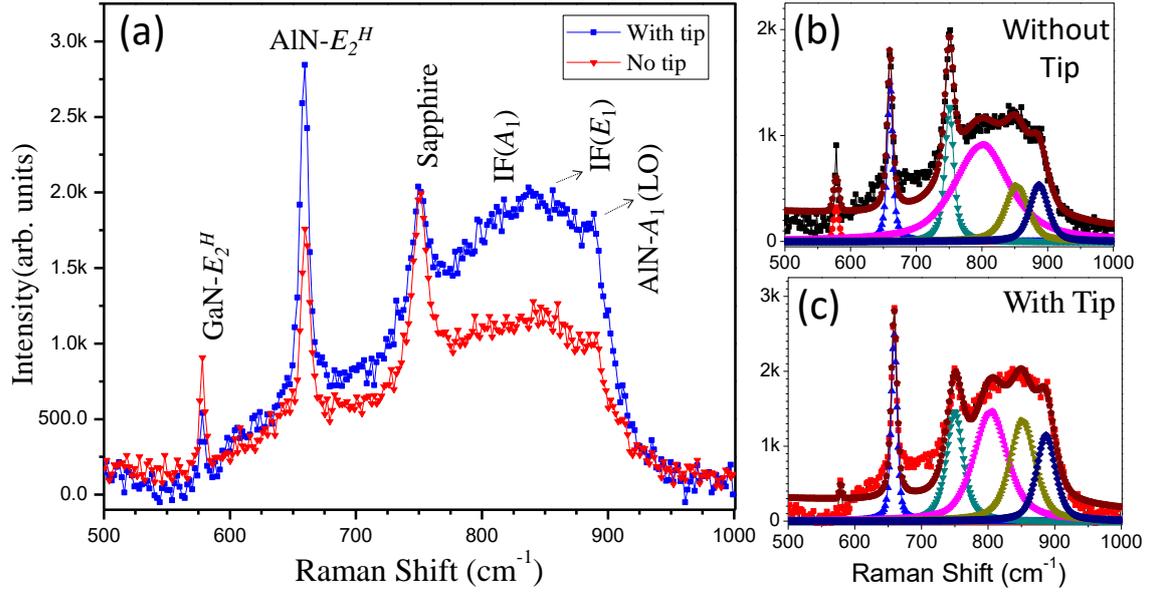

FIG. S2. a) Far- and near-field Raman spectra recorded at 300K for AlGaN MQWs without and with the presence of TERS probe, respectively. Deconvoluted spectra correspond to Raman spectra b) without (far-field) and c) with (near-field) TERS probe.

Table S1. Parameters extracted from fitted Raman spectra without and with TERS tip

| Modes | Intensity (Without tip) | Intensity (With tip) | Enhancement factor |
|---|---|---|---|
| AlN-$E_2^H$ | 1526 | 2498 | 788 |
| IF($A_1$) | 915 | 1462 | 740 |
| IF($E_1$) | 526 | 1343 | 1922 |

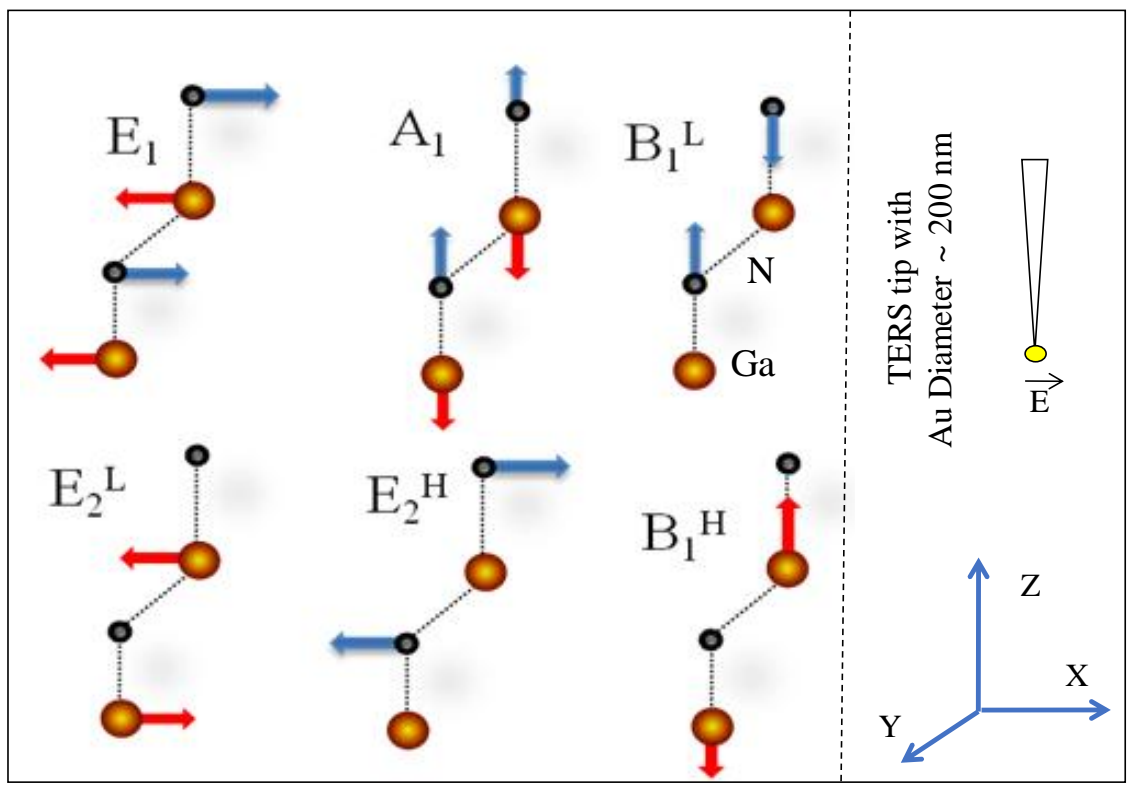

FIG. S3. Different symmetry allowed vibrational modes in wurtzite crystal and their activities along crystalline axes shown in the outset. Direction for electric field of the light (electro-magnetic wave) excitation with respect to TERS tip is also shown in the outset.